\documentclass[a4paper,11pt]{article}
\pdfoutput=1 

\usepackage{jheppub} 

\usepackage[T1]{fontenc} 
\usepackage[utf8]{inputenc}
\usepackage{amsfonts}
\usepackage{amsbsy}
\usepackage{url}
\usepackage{enumerate}
\usepackage[justification=raggedright]{caption}
\usepackage{hhline}
\usepackage{graphicx,color}
\usepackage{amssymb,amsmath}
\usepackage{slashed}
\usepackage{hyperref}
\usepackage{braket}
\usepackage{simplewick}
\usepackage{ascmac}
\usepackage{latexsym}
\usepackage{pifont}          
\usepackage{bm}
\usepackage{comment}

\newcommand{\h}[1]{\hspace{#1}}
\newcommand{\mr}[1]{\mathrm{#1}}

\title{\boldmath Electroweak-Skyrmion as Asymmetric Dark Matter}
\preprint{KEK-TH-2345}

\author[a]{Yu Hamada,}
\author[a,b]{Ryuichiro Kitano,}
\author[c]{and Masafumi Kurachi}
\affiliation[a]{Theory Center, High Energy Accelerator Research Organization (KEK), Tsukuba 305-0801, Japan}
\affiliation[b]{The Graduate University for Advanced Studies (Sokendai), Tsukuba 305-0801, Japan}
\affiliation[c]{Research and Education Center for Natural Sciences, Keio University,  Yokohama 223-8521, Japan}

\emailAdd{yuhamada@post.kek.jp}
\emailAdd{ryuichiro.kitano@kek.jp}
\emailAdd{kurachi@keio.jp}

\abstract{
We propose a scenario that the Electroweak-Skyrmion, a solitonic object made of the Higgs field and the electroweak gauge fields, 
is identified as an asymmetric dark matter. 
In this scenario, the relic abundance of the dark matter is related to the baryon asymmetry of the Universe through a sphaleron-like process. 
We show that the observed ratio of dark matter abundance to the baryon asymmetry can be explained by this scenario 
with an appropriate choice of model parameters that is allowed by currently available experimental constraints.
}

\begin{document} 
\maketitle
\flushbottom

\section{Introduction}

It is observationally known that the baryon and the dark mater (DM) abundances in the Universe are close to each other: $\rho_{\rm DM}/\rho_{\rm B}\simeq 5$. 
This fact tempts us to pursue the possibility that these abundances have a common origin. 
Among various DM models, asymmetric DM (ADM) scenario 
\cite{Nussinov:1985xr,Harvey:1990qw,Barr:1990ca,Barr:1991qn,Dodelson:1991iv,Kaplan:1991ah,Kuzmin:1996he,Foot:2003jt,Kitano:2004sv,Gudnason:2006ug,Kitano:2008tk,Kaplan:2009ag} 
(see Refs.~\cite{Davoudiasl:2012uw,Petraki:2013wwa,Zurek:2013wia} for reviews) 
is one of those 
that have a potential ability to address the similarity of DM and baryon asymmetries. 
In the ADM  scenario, the current DM abundance is described as DM particle-antiparticle asymmetry, 
which has a common origin with the baryon asymmetry.

In Refs.~\cite{Kitano:2016ooc,Kitano:2017zqw}, it was shown that 
a solitonic object made of the Higgs field is topologically stable 
with a minimal addition of higher-derivative operators to the Standard Model (SM) Higgs Lagrangian 
as long as the electroweak (EW) gauge fields are ignored, 
and can be a DM candidate. 
Since the mathematical structure of the object is analogous to the Skyrmion solution in the effective theory of QCD, 
it was called the Electroweak-Skyrmion (EW-Skyrmion) \cite{Eilam:1985tg,Brihaye:1989kz,Carson:1990yk} 
(see Ref.~\cite{Criado:2020zwu} for a recent work). 
The treatment of ignoring the gauge interactions is justified 
when the product of the gauge coupling constant squared and the coefficient of the higher dimensional operator is small enough.
Therefore, the existence of the Skyrmion solution with the Higgs boson mass, $125\, \mr{GeV}$, has been established.
It is however important to consider the effects of the gauge interaction for the discussion of very heavy EW-Skyrmions
as well as of the high temperature physics in cosmology.

In this paper, we incorporate this idea of the EW-Skyrmion as a DM into the ADM scenario.
It has been shown in Ref.~\cite{Kitano:2016ooc} that 
the thermal relic abundance is too small to explain the amount of DM for TeV scale Skyrmions 
due to too large annihilation cross sections.
Thus there must exist asymmetry between the abundances of the EW-Skyrmion and anti-EW-Skyrmion in the universe to avoid this annihilation.
Actually, when the $SU(2)_W$ gauge fields are taken into account, 
the EW-Skyrmion is not topologically stable and the Skyrmion number (DM number) is not conserved.
It is meta-stable at best, that is, a local minimum of the energy functional 
when the effect of the gauge coupling constant is small,
and can decay by quantum tunneling or thermal processes.
Because the decay process produces the $B+L$ number via the $B+L$ anomaly,
we can relate the baryon asymmetry and the DM number relic abundance in the early universe.
This scenario is similar to that in Ref.~\cite{Barr:1990ca},
in which the technibaryon plays a role of the ADM.
Interestingly, the stability and the mass of the EW-Skyrmion strongly depends on the magnitudes of the higher-derivative terms,
which can be directly measured by observing scattering process of the EW gauge bosons. 
Thus our ADM scenario is testable by not only DM direct detection experiments such as XENON experiment
but also collider experiments
with synergetic effects.
We present the current experimental constraints on our model.

This paper is organized as follows.
Sec.~\ref{sec:Sky} is devoted to a demonstration that the EW-Skyrmion solution exists even with the gauge fields included.
In Secs.~\ref{sec:model} and \ref{sec:without-gauge}, 
we review the model given in Ref.~\cite{Kitano:2016ooc} and the EW-Skyrmion without the gauge fields.
In Sec.~\ref{sec:gauge} we incorporate the EW gauge interaction into the model 
and discuss its impact to the stability of the EW-Skyrmion. 
In Sec.~\ref{sec:abundance}, we discuss how the asymmetry of EW-Skyrmion density is related to the baryon asymmetry 
through the (inverse) decay process, 
and show that observed ratio of dark matter abundance to the baryon asymmetry can be explained by this scenario 
with an appropriate choice of model parameter that is allowed by currently available experimental constraints.
We summarize the discussion in Sec.~\ref{sec:summary}.

\section{Electroweak-Skyrmion}\label{sec:Sky}
\subsection{The Higgs Lagrangian}
\label{sec:model}
We start from introducing higher derivative terms into the SM Higgs Lagrangian:
\begin{equation}
{\cal L} = {\cal L}_{\rm SM} + {\cal L}_{p^4}.
\label{eq:Lagrangian}
\end{equation}
${\cal L}_{\rm SM}$ is the Lagrangian of the Higgs sector of the SM, which is defined in the following way:
\begin{equation}
{\cal L}_{\rm SM} =  \frac{v_{\rm EW}^2}{4} \left( 1 + \frac{h(x)}{v_{\rm EW}} \right)^2 {\rm Tr}\left[  D_\mu U(x)\,  D^\mu U(x)^\dagger  \right]      \,+\, \frac{1}{2}\partial_\mu h(x)\partial^\mu h(x) - V(h(x)).
\label{eq:LSM}
\end{equation}
Here, $v_{\rm EW} (\simeq 246 {\rm GeV})$ is the vacuum expectation value (VEV) of the Higgs field and we defined the Higgs $H(x)$ field by a two-by-two matrix notation with the ``polar decomposition'' as
\begin{equation}
 H(x) = \left(1+\frac{h(x)}{v_{\rm EW}} \right) U(x),
\end{equation}
where $h(x)$ is the field which represents the physical Higgs boson and $U(x)$ represents the Nambu-Goldston (NG) degree of freedom of the Higgs field:
\begin{equation}
 U(x) = e^{i\, \pi^i(x)\, \sigma^i/v_{\rm EW} }\ \ \ \ \left(\sigma^i : {\rm Pauli\ matrix}\right).
\end{equation}
The fields $h(x)$ and $U(x)$ defined in this way have transformation properties under 
a $SU(2)_L \times SU(2)_R$ global transformation as:
\begin{align}
h(x) &\longrightarrow h(x),\\
U(x) &\longrightarrow L \,U(x)\, R^\dagger,
\end{align}
where $L$ and $R$ represents the elements of $SU(2)_L$ and $SU(2)_R$, respectively. The covariant derivative of $U$ thus is expressed as:
\begin{equation}
D_\mu U(x) \equiv \partial_\mu U(x) -i g W_\mu^a \left(\frac{\sigma^a}{2}\right) U(x) + i g' U(x) Y_\mu \left(\frac{\sigma^3}{2}\right),
\end{equation}
where $W_\mu^a$ and $Y_\mu$ are the weak and the hypercharge gauge fields, and $g$ and $g'$ are their coupling constants, respectively.
The potential term $V(h(x))$ appearing in Eq.~\eqref{eq:Lagrangian} takes the form that is same as the SM:
\begin{equation}
V(h(x)) = \lambda v_{\rm EW}^2\, h(x)^2 +  \lambda v_{\rm EW}\, h(x)^3 + \frac{\lambda}{4}\, h(x)^4.
\label{eq:Vh}
\end{equation}
The higher derivative part ${\cal L}_{p^4}$ in Eq.~(\ref{eq:Lagrangian}), 
which we call ``Skyrme term'' throughout the paper, is defined as:
\begin{align}
{\cal L}_{p^4} &=  \alpha_4 {\rm Tr}\left[  D_\mu U(x)^\dagger\,  D_\nu U(x)  \right] {\rm Tr}\left[  D^\mu U(x)^\dagger\,  D^\nu U(x)  \right] \nonumber \\
& +\, \alpha_5 \left({\rm Tr}\left[ D_\mu U(x)^\dagger\,  D^\mu U(x)  \right] \right)^2 .
\label{eq:Lp4}
\end{align}

Note that the Lagrangian \eqref{eq:Lagrangian} is invariant under the global $SU(2)_L \times SU(2)_R$ transformation 
when the $U(1)_Y$ coupling is ignored.
The VEV of the Higgs field $H(x)$, $h=\pi^i=0$, 
breaks this symmetry as
\begin{equation}
 SU(2)_L \times SU(2)_R \longrightarrow SU(2)_V ,\label{210328_1Jul21}
\end{equation}
where $SU(2)_V$ is defined as a diagonal subgroup of $SU(2)_L \times SU(2)_R$ and called the custodial symmetry.

In the limit of $\alpha_4, \alpha_5 \to 0$, the Lagrangian defined above is nothing but the SM Higgs Lagrangian. 
The newly added terms (Skyrme terms) describe anomalous quartic gauge interactions among EW gauge bosons. 
Phenomenologically, we know only the upper bound of the coefficients $\alpha_4$ and $\alpha_5$ 
from the experiments of the EW gauge boson scatterings. 
We will present such experimental bounds later.
As a possible origin of the terms, the exchange of heavy resonances generate the $\mathcal{O}(p^4)$ terms in the effective theory.

In the view point of an effective field theory, $h(x)$ and $U(x)$ do not have to form a single linear field in general, 
thus the factor in front of the first term of the Lagrangian (\ref{eq:Lagrangian}) could be any function of $h(x)/f$ with $f$ being a scale which is not necessarily related to $v_{\rm EW}$. Also a similar factor can be multiplied to the Skyrme term as well. There could be other types of higher order terms in the Lagrangian, and the form of $V(h(x))$ does not have to be limited to the one shown in Eq.~(\ref{eq:Vh}) as well. The choice of the Lagrangian above is to make the study tractable and to make the difference from the SM Lagrangian as minimal as possible: only the difference between the Lagrangian discussed in this section and that of the SM is the existence of the $\mathcal{O}(p^4)$ terms. 
There are other $\mathcal{O}(p^4)$ terms that involve the field strength of the $SU(2)$ gauge fields.
We assume that the coefficients of those terms are small enough such that the qualitative discussion remains the same.

\subsection{EW-Skyrmion without gauge fields (review)}
\label{sec:without-gauge}
Because the symmetry breaking pattern in Eq.~\eqref{210328_1Jul21} is similar to 
that of the chiral symmetry breaking in two-flavor QCD,
we can expect an existence of a Skyrmion-like soliton described by the Higgs field,
which is called the EW-Skyrmion.
In particular, they are exactly the same
when the radial component scalar $h(x)$ and the gauge fields are absent.

It is known that the existence of the EW-Skyrmion is topologically ensured 
even when $h(x)$ is included, as long as the gauge fields are ignored.
To see this, we set $g=g'=0$ in the covariant derivatives for a while.
For simplicity, we take $\alpha_4=-\alpha_5\equiv \alpha$.
This model is same as the one discussed in Refs.~\cite{Kitano:2016ooc,Kitano:2017zqw}. 
Then the Lagrangian becomes
\begin{align}
 \cal L =&  \frac{v_{\rm EW}^2}{4} \left( 1 + \frac{h(x)}{v_{\rm EW}} \right)^2 {\rm Tr}\left[  \partial_\mu U(x)\,  \partial^\mu U(x)^\dagger  \right]      \,+\, \frac{1}{2}\partial_\mu h(x)\partial^\mu h(x) - V(h(x))\nonumber \\
           & + \frac{1}{2}\,\alpha\, {\rm Tr} \left[\partial_\mu U(x)\, U(x)^\dagger \, ,\, \partial_\nu U(x)\, U(x)^\dagger\right]^2,
\label{eq:L}
\end{align}
where $[A, B]\equiv AB-BA $. 
%

To describe the EW-Skyrmion, we take the hedgehog ansatz for the static configuration of $U(x)$ as 
\begin{equation}
 U(x)=e^{i \theta(r) \sigma^i \hat x_i},\label{144403_10May21}
\end{equation}
where
\begin{equation}
 r \equiv \sqrt{x_i x_i}, \h{2em} \hat x_i \equiv x_i /r.
\end{equation}
As for $h(x)$, we assume that the static solution, $h_0(x)$, is spherically symmetric:
\begin{equation}
 h_0(x)/v_{\rm EW} = \phi(r).\label{201903_3Jul21}
\end{equation}

\begin{figure}[tbp]
\centering
\includegraphics[width=0.6\textwidth]{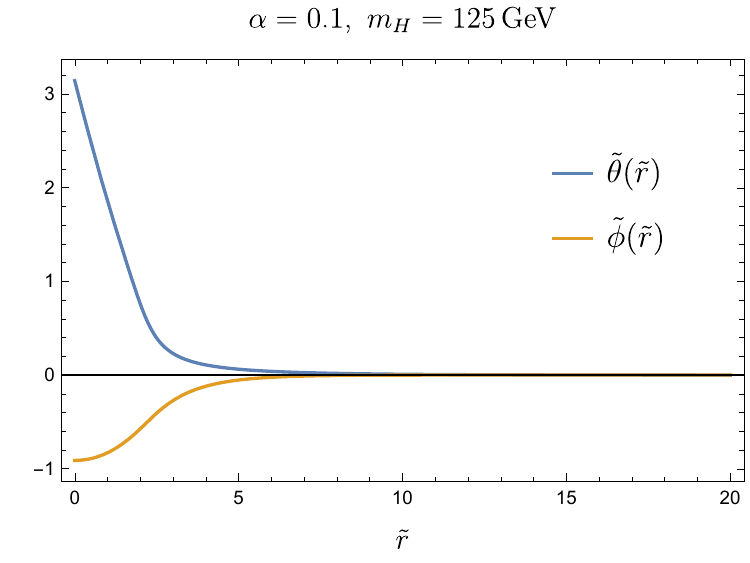}
\caption{Plots of the $r$-dependence of the profile functions $\theta, \phi$. 
We have taken $\alpha_4 = -\alpha_5 (\equiv\alpha)=0.1$
and $m_H = 125 \, \mr{GeV}$.
The energy of this configuration is calculated as $14.8\, \mr{TeV}$.
}
\label{003358_12Jul21}
\end{figure}

\begin{figure}[tbp]
\centering
\includegraphics[width=0.6\textwidth]{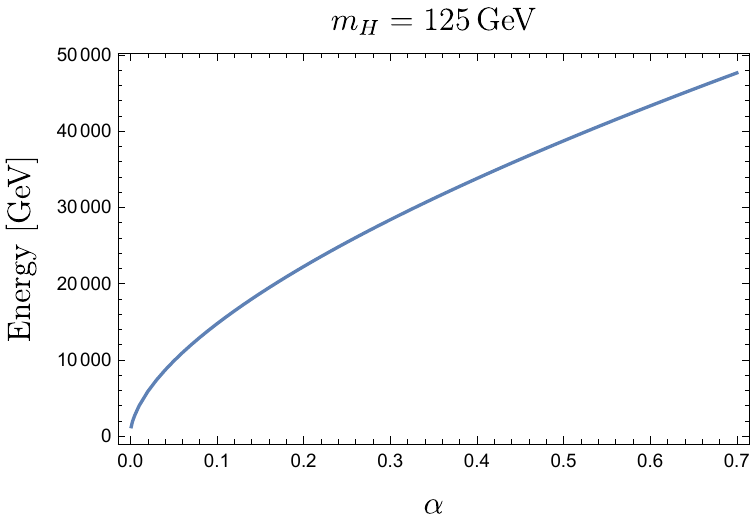}
\caption{The EW-Skyrmion energy as a function of $\alpha$.
We have taken $\alpha_4 = -\alpha_5 \equiv\alpha$
and $m_H = 125 \, \mr{GeV}, v_\mr{EW} = 246 \, \mr{GeV}$.
The energy approaches to $0$ as $\alpha\to 0$.
}
\label{010740_12Jul21}
\end{figure}

With these ansatz, the total energy of the system takes the following form:
\begin{align}
E\left[\tilde{\theta}(\tilde{r}), \tilde{\phi}(\tilde{r})\right] = 2\pi \left( \frac{v_{\rm EW}}{e}\right) \int_0^\infty &d\tilde{r}\tilde{r}^2\, \Bigg[
 \  \left( 1 + \tilde{\phi}(\tilde{r}) \right)^2 \left( \tilde{\theta}'(\tilde{r})^2 + 2\, \frac{ \sin^2\tilde{\theta}(\tilde{r}) }{\tilde{r}^2} \right)
 \nonumber\\
&\ +\ \frac{ \sin^2\tilde{\theta}(\tilde{r}) }{\tilde{r}^2} \left( \frac{ \sin^2\tilde{\theta}(\tilde{r}) }{\tilde{r}^2} + 2 \tilde{\theta}'(\tilde{r})^2 \right)
 \nonumber\\
 &\ +\  \tilde{\phi}'(\tilde{r})^2 \ +\  \frac{\,m_h^2\,}{e^2\,v_{\rm EW}^2}\left(  \tilde{\phi}(\tilde{r})^2+\tilde{\phi}(\tilde{r})^3+\frac{1}{4}\tilde{\phi}(\tilde{r})^4\right) \Bigg].\nonumber \\
\label{eq:E2}
\end{align}
%
Here we have defined ``Skyrme coupling''
$e \equiv 1/(4\sqrt{\alpha})$
and the dimensionless variable
\begin{equation}
  \tilde r \equiv \frac{r \, v_\mr{EW}}{4 \sqrt{\alpha}} \, ,
\end{equation}
and  $\tilde{\phi}(\tilde{r}) \equiv \phi(r) $.
We have replaced the parameter $\lambda$ by the mass of the scalar $m_h$ by using the relation $\lambda=m_h^2/(2v_{\rm EW}^2)$. The energy is minimized when $\tilde{\theta}(\tilde{r})$ and $\tilde{\phi}(\tilde{r})$ satisfy the following coupled equations:
\begin{align}
\left(1+\tilde{\phi}(\tilde{r})\right)^2 \left( -\sin 2\tilde{\theta}(\tilde{r})  + 2 \tilde{r}\tilde{\theta}'(\tilde{r}) + \tilde{r}^2\tilde{\theta}''(\tilde{r})\right)
+ 2 \left(1+\tilde{\phi}(\tilde{r})\right) \tilde{\phi}'(\tilde{r})\, \tilde{r}^2\tilde{\theta}'(\tilde{r})  &  \nonumber\\
- \frac{ \sin^2\tilde{\theta}(\tilde{r}) \sin2\tilde{\theta}(\tilde{r}) }{\tilde{r}^2} + \sin2\tilde{\theta}(\tilde{r})\,\tilde{\theta}'(\tilde{r}) ^2  + 2\sin^2\tilde{\theta}(\tilde{r})\,\tilde{\theta}''(\tilde{r})
= &0.\ \ 
\label{eq:Eq2}
\end{align}
\begin{align}
\left(1+\tilde{\phi}(\tilde{r})\right)\left(  \tilde{r}^2\tilde{\theta}'(\tilde{r}) + 2 \sin^2\tilde{\theta}(\tilde{r}) \right)
-2\tilde{r} \tilde{\phi}'(\tilde{r})-\tilde{r}^2 \tilde{\phi}''(\tilde{r})& \nonumber\\
+\, \frac{1}{2}\,\frac{m_h^2}{\, e^2 \, v_{\rm EW}^2\,}\, \tilde{r}^2 \left( 2\,\tilde{\phi}(\tilde{r})  + 3\, \tilde{\phi}(\tilde{r})^2 + \tilde{\phi}(\tilde{r})^4 \right) &= 0.
\label{eq:Eq3}
\end{align}
We numerically solve the above equations with the following boundary conditions
\begin{equation}
\tilde \phi'(0) = 0, \h{2em} \tilde \theta(0) = \pi\label{163835_12Jul21}
\end{equation}
\begin{equation}
\tilde \phi(\tilde r) \to 0,\h{1em}  \tilde \theta(\tilde r) \to 0 \h{1em}  (\tilde r \to \infty) \, ,\label{163850_12Jul21}
\end{equation}
and find that the system always has topologically a non-trivial field configuration as far as $\alpha >0$,
which is identified with the EW-Skyrmion. 
In Fig.~\ref{003358_12Jul21}, we show the example of $\tilde{\theta }(\tilde{r})$ (upper blue curve) and $\tilde{\phi}(\tilde{r})$ (lower orange curve) in the case of $\alpha=0.1 (e\simeq 0.79)$, $v_\mr{EW}=246$ GeV, and $m_h = 125$ GeV.
Fig.~\ref{010740_12Jul21} shows how the mass of the EW-Skyrmion depends on the input value of $\alpha$.

Let us discuss the stability of the EW-Skyrmion.
As the original Skyrmion in the Skyrme model,
it is topologically stable
because the configuration has a non-trivial winding number (Skyrmion number) $N_{H}$
defined as
\begin{equation}
 N_{H} \equiv -\frac{\epsilon_{ijk}}{24 \pi^2} \int d^3 x \, \mr{Tr}\left[V_i V_j V_k\right] \, ,\label{151605_10May21}
\end{equation}
with $V_\mu \equiv (\partial_\mu U )U^\dagger$.
Under the compactification of the three-dimensional space $\mathbb{R}^3$ to $S^3$,
this quantity always amounts to an integer value.
Thus all possible configurations are classified into topological sectors labeled with this integer.
This follows from a mathematical fact that the vacuum manifold
\begin{equation}
\mathcal{M} = \frac{SU(2)_L \times SU(2)_R}{SU(2)_V} \simeq S^3 
\end{equation}
has a non-trivial third homotopy group $\pi_3(\mathcal{M}) = \mathbb{Z}$.
The field $U(x)$ maps the three dimensional space $\mathbb{R}^3 \cup \{\infty\} \simeq S^3$ into the vacuum manifold $S^3$,
where $\{\infty\}$ is the point at infinity,
and hence $N_H$ is nothing but the winding number corresponding to this map
and is a topological invariant.
The configuration of the EW-Skyrmion, Eq.~\eqref{144403_10May21}, with the boundary condition Eqs.~\eqref{163835_12Jul21} and \eqref{163850_12Jul21} has $N_H=1$, and 
cannot be continuously deformed to the vacuum in which the EW-Skyrmion is absent ($N_H=0$).
Therefore, the EW-Skyrmion is topologically stable 
as far as $g=g'=0$ and $\alpha >0$.
For $\alpha <0$, there is no stable solution with non-trivial winding numbers.

We assume that the EW-Skyrmion is electrically neutral.
The quantum numbers depend on the UV theory to induce the Skyrme terms,
and are in general encoded in the Wess-Zumino-Witten term~\cite{Adkins:1983ya}.
Here we assume that the UV theory does not induce the WZW term so that the EW-Skyrmion is neutral and quantized as a spin-0 state.
Examples include an $SU(2)$ gauge theory with two flavors of quarks, 
which can give a neutral and spin-0 baryon, while we do not specify the UV model in this paper.

\subsection{EW-Skyrmion coupled with gauge fields}
\label{sec:gauge}
We here discuss effects of the $SU(2)_W$ gauge interactions.
For simplicity, throughout this paper, we ignore the $U(1)_Y$ gauge field, \textit{i.e.}, $g'=0$,
which is expected not to affect the present argument drastically
because the effect of $g'\neq 0$ is quite small in the case of the electroweak sphaleron~\cite{Manton:1983nd,Klinkhamer:1984di}.

Firstly, we should note that the solution for $F(r)$ and $\phi(r)$ obtained above is no longer consistent with 
$W_i^a=0$ when $g\neq 0$,
because the $SU(2)_W$ current, which is given by
\begin{align}
J_i^{a}& \equiv \left. \frac{\delta \mathcal{L}}{\delta W_i^a} \right|_{W=0} \\
 &=  2ig \frac{v_{EW}^2}{4}\left(1+\frac{h(x)}{v_{EW}}\right)^2
\mr{Tr}\left[\frac{\sigma^a}{2}V_i\right]
- 2ig\, \alpha \, \mr{Tr}\left(\left[\frac{\sigma^a}{2},V_\nu \right]\left[V_i , V^\nu\right]\right) \, , \label{144500_24May21}
\end{align}
does not vanish in general. (We have used $V_\mu \equiv (\partial_\mu U )U^\dagger$.)
The induced amount of the gauge field $W_i^a$ can be estimated
based on the perturbation theory with respect to $g$
through the linearized EOM:
\begin{equation}
\partial_j (\partial_j W_i^{a} - \partial_i W_j^{a}) = -J_i^{ a} \, \label{111931_24May21}
\end{equation}
\begin{equation}
 \therefore \partial ^2 W \sim g v_\mathrm{EW} ^2 R ^{-1} + g \alpha R^{-3}\label{144030_24May21}
\end{equation}
where $R \sim \sqrt{\alpha} (v_\mathrm{EW})^{-1}$ is the typical size of the EW-Skyrmion for $g=0$, and we have omitted the indices $a$ and $i$ for a rough estimation.
The typical size is understood by the condition
\begin{equation}
 v_\mathrm{EW}^2 R^{-2} \sim \alpha R^{-4} ,
\end{equation}
where the left-hand side and the right-hand side represent the $\mathcal{O}(p^2)$ and  $\mathcal{O}(p^4)$ terms in the Lagrangian, respectively,
and are balanced to obtain a stable solution.
Thus the induced $W$ is rewritten as
\begin{equation}
 W \sim g \left(v_\mathrm{EW}^2 R + \alpha R^{-1}\right)  \sim g \sqrt{\alpha} \, v_\mathrm{EW} .
\end{equation}
Accordingly, the energy is shifted from the original one by
\begin{align}
 \Delta E &= \int d^3 x \left[\frac{1}{4}W_{ij}^a W_{ij}^a - W_i^a J_i^{a}\right] 
= -\frac{1}{2}\int d^3 x W_i^a J_i^{a} \\
&\sim - R^3 \times g\sqrt{\alpha} v_\mr{EW} \times g \alpha R^{-3}
 \sim - g^2 \alpha^{3/2} v_\mr{EW}\, .
\end{align}
Compared with the original energy of the order of $\mathcal{O}(\sqrt{\alpha} \, v_\mr{EW})$,
the original solution is deformed by the order of $g^2 \alpha$, 
instead of $g^2$ alone.
Therefore, as long as $g^2 \alpha\ll 1$,
the gauge interaction can be thought of as a small perturbation to the discussion in the previous section.
For $g^2 \alpha\gtrsim 1$, the solution is destroyed as we see later.

In such a deformed configuration with $W \neq 0$,
the topological stability is not ensured any more.
This is understood by noticing that there is no gauge-invariant topological charge.
Indeed, the Skyrmion number \eqref{151605_10May21} is not gauge invariant,
and its winding can be removed by a large gauge transformation.
An alternative gauge invariant quantity labeling the configuration is
\begin{equation}
Q \equiv N_H + N_{CS} \, ,\label{173008_4Jul21}
\end{equation}
where $N_{CS}$ is the Chern-Simons number, 
\begin{equation}
 N_{CS} \equiv \frac{g^2}{16 \pi^2}\int d^3 x \, \epsilon_{ijk} \mathrm{Tr}\left[W_{ij} W_k + \frac{2ig}{3}W_i W_j W_k\right] \, .
\end{equation}
Although the individual quantities $N_H$ and $N_{CS}$ are not gauge invariant, $Q$ is so
because the large-gauge dependence in $N_H$ is compensated by that from $N_{CS}$.
However, $Q$ is not topologically conserved nor an integer value.
Instead, $Q$ can be continuously changed as
\begin{equation}
 \Delta Q = \Delta N_{CS} = \frac{g^2}{16 \pi^2}\int d^4 x \, \mr{Tr}[F \tilde F] 
\end{equation}
and the right hand side can take an arbitrary value in general.
(Note that $N_H$ is topologically conserved, $\Delta N_H=0$.)
Thus any static configuration with $Q\neq 0$ can be continuously deformed into the vacuum corresponding to $Q=0$.
Therefore, the EW-skyrmion with non-zero $g$ cannot be topologically stable, but ``classically stable'' at best.
The decay rate due to the tunneling is estimated to be $\exp(-8\pi^2/g^2)$,
which is small enough as in the case of proton decay 
through the instanton effect.

The (classically) stable EW-Skyrmion is indeed found for small $g^2 \alpha$.
To obtain a solution of the EOMs
beyond the perturbation theory with respect to $g^2 \alpha$,
we have to rely on a numerical analysis.
To do so, let us construct an ansatz of the gauge fields $W^a_i$
in addition to the ansatz of the Higgs, $U(x)$ and $\phi(x)$, Eqs.~\eqref{144403_10May21} and \eqref{201903_3Jul21}.
A spherically symmetric ansatz for $W_i$ is given as \cite{Eilam:1985tg,Ratra:1987dp}
\begin{equation}
g W_i^a = \frac{\mathrm{Re}\chi(r) -1}{r} \epsilon_{iab} \hat{x}_b 
- \frac{\mathrm{Im}\chi(r) }{r} (\delta_{ia}- \hat{x}_i \hat{x}_a)
- \delta(r) \hat{x}_i \hat{x}_a\label{194727_17Jul21}
\end{equation}
and $W_0^a=0$.

We should note that these ansatz have a residual gauge symmetry generated by a gauge transformation
\begin{equation}
\mathcal{G} = \exp \left[\omega(r) \hat{x}_i \frac{\sigma_i}{2}\right] \, ,
\end{equation}
which acts as
\begin{align}
 \chi(r) &\to e^{i \omega(r)} \chi (r) \\
 \theta(r) &\to \theta(r) + \frac{\omega(r)}{2}  \\
 \delta(r) &\to \delta(r) + \partial_r \omega(r) \, .
\end{align}
To fix this redundancy, we take a gauge such that $\mathrm{Im}\chi(r)=0$.

After substituting the above ansatz, 
we obtain the energy functional as

\begin{align}
 \frac{E}{4\pi}&= \int _0 ^\infty dr ~ \Big[ \frac{1}{2 g^2 r^2}
\left[2(r^2 \delta^2 -1) \chi^2 + 2 r^2 \chi'^2 +\chi^4 +1\right] \nonumber\label{185321_22May21} \\
&+ \frac{1}{8}(\delta^2 - 4 \delta\theta')\left[r^2 v^2 (1+\phi)^2 + 8 \alpha (1+\chi^2 -2 \chi \cos 2\theta)\right] \nonumber \\
&+ \frac{1}{4} \lambda r^2 v^4 (\phi^4 + 4 \phi^3 + 4 \phi^2)
+ \frac{1}{2}r^2 v^2 (\theta'^2 + \phi'^2)
+\frac{1}{4} (v^2 + 16 \alpha\theta'^2)(1+\chi^2 -2 \chi \cos 2\theta)  \nonumber\\
&+\frac{1}{2 r^2}\alpha (1+\chi^4 -4 \chi^3 \cos 2\theta -4\chi \cos 2\theta + 2\chi^2\cos4\theta + 4\chi^2 ) \nonumber \\
&+ \frac{1}{4}v^2 (\phi^2 + 2\phi) (2r^2 \theta'^2 - 2\chi \cos 2\theta + \chi^2 +1) \Big] \, .
\end{align}
where we have assumed $\alpha_4=-\alpha_5=\alpha$ for simplicity
and the prime denotes the derivative with respect to $r$.
Note that $\delta$ does not have a kinetic term, which means that it is an auxiliary field and is explicitly solvable.
The solved value is 
\begin{equation}
\delta = 
2 \theta' 
\left(1- \frac{8 \chi^2}{g^2}
\left[r^2 v^2 (1+\phi)^2 + 8 \alpha (1 + \chi^2 -2 \chi \cos 2\theta) + 8 \chi ^2 /g^2\right]^{-1}
\right) 
\equiv \bar{\delta}.\label{191306_24May21}
\end{equation}

\begin{figure}[tbp]
\centering
\includegraphics[width=0.48\textwidth]{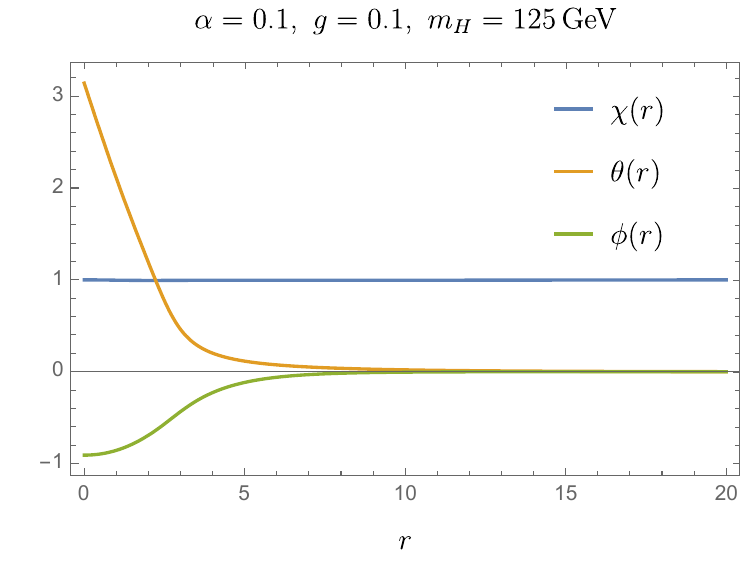} \hspace{0.5em}
\includegraphics[width=0.48\textwidth]{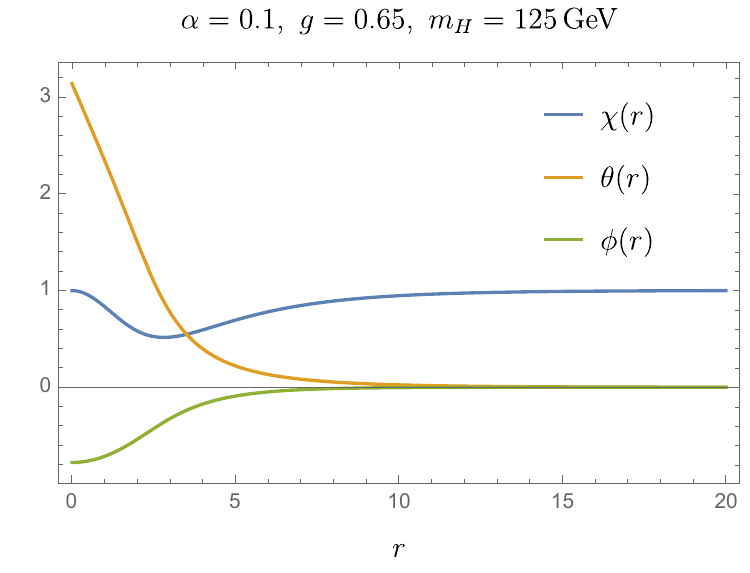} \\[2ex]
\includegraphics[width=0.48\textwidth]{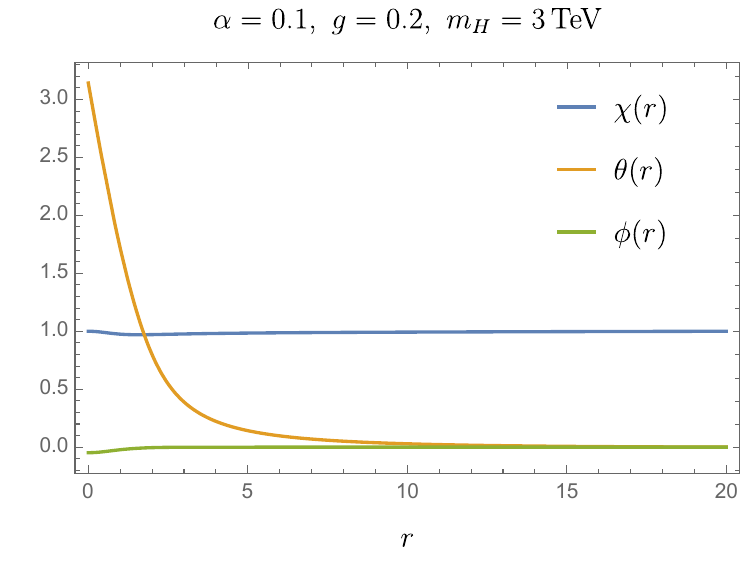}
\caption{
Plots of the $r$-dependence of the profile functions $\chi, \theta, \phi$. 
We have taken $\alpha_4 = -\alpha_5 \equiv\alpha = 0.1$.
The length unit is taken so that $v_\mr{EW}=1$.
The physical point of $g$ and $m_H$ corresponds to the upper right panel.
$Q=N_H+N_{CS}$ is obtained as $1-1.9\times10^{-5}$, $1-0.054$, and $1-6.5\times10^{-4}$ in the upper-left, upper-right, and lower panels, respectively.
}
\label{203149_3Jul21}
\end{figure}

\begin{figure}[tbp]
\centering
\includegraphics[width=0.47\textwidth]{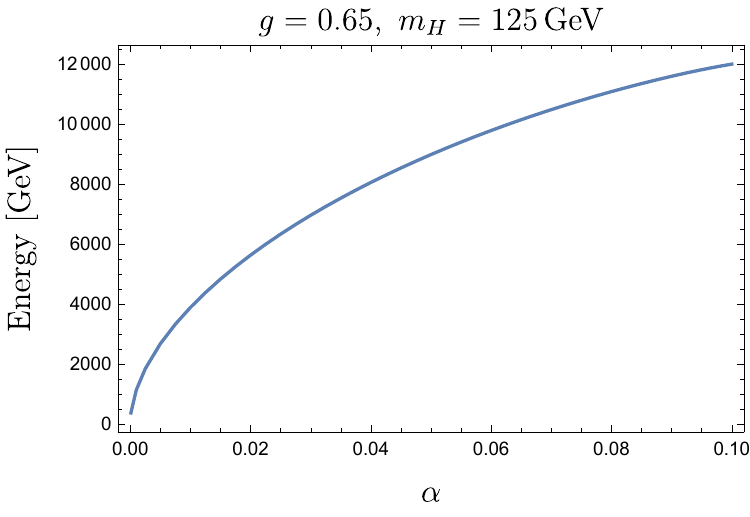} \hspace{0.5em}
\includegraphics[width=0.47\textwidth]{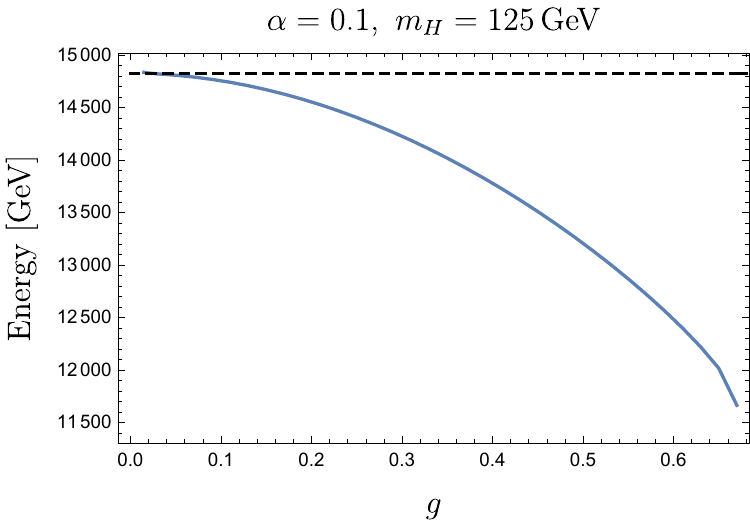}
\caption{Mass of the EW-Skyrmion 
versus $\alpha$ ($g$) in the left (right) panel, respectively.
The black dashed line in the right panel indicates the energy calculated by the ansatz without the gauge field (Fig.~\ref{003358_12Jul21}).
No stable solutions are found for $\alpha > 0.11$ in the left panel 
and $g > 0.67$ in the right panel.
}
\label{203318_3Jul21}
\end{figure}

Under the constraint \eqref{191306_24May21},
we solve the EOMs for $\chi$, $\phi$, and $\theta$ numerically.
We have imposed a boundary condition for $\chi$ as
\begin{equation}
 \chi(0) = \chi (\infty) =1 
\end{equation}
for regularity and finiteness of the energy.
The boundary conditions for $\phi$ and $\theta$ are the same as the previous ones, 
Eqs.~\eqref{163835_12Jul21} and \eqref{163850_12Jul21}.
In Fig.~\ref{203149_3Jul21}, the $r$-dependence of the solutions are shown 
for several parameter values keeping $\alpha =0.1$.
The qualitative behaviors of $\theta$ and $\phi$ are not different from those with $g=0$.
Fig.~\ref{203318_3Jul21} shows that the relation between the energy (mass) of the EW-Skyrmion and the parameters $g, \alpha$.
Note that, no stable solution is found for $g^2 \alpha \gtrsim (0.21)^2$.
This is consistent with a perturbative analysis (see Appendix \ref{App:instability}),
which predicts an emergence of instability for large $g^2 \alpha$.

In addition, the gauge invariant quantity, $Q=N_H+N_{CS}$, is indeed non-integer values for the obtained solutions because $N_{CS}$ is not an integer in general.
In Fig.~\ref{203149_3Jul21}, for instance, $Q$ is calculated as $1-1.9\times10^{-5}$, $1-0.054$, and $1-6.5\times10^{-4}$ in the upper-left, upper-right, and lower panels, respectively.
It deviates from unity by at most $-0.22$ for $ g=0.65$ and $\alpha=0.109$.
This deviation is not due to the numerical error but is consistent with previous studies in the literature, 
\textit{e.g.}, Ref.~\cite{Eilam:1985tg}, 
in which the light scalar field (physical Higgs boson) is absent.

\begin{figure}[tbp]
\centering
\includegraphics[width=0.49\textwidth]{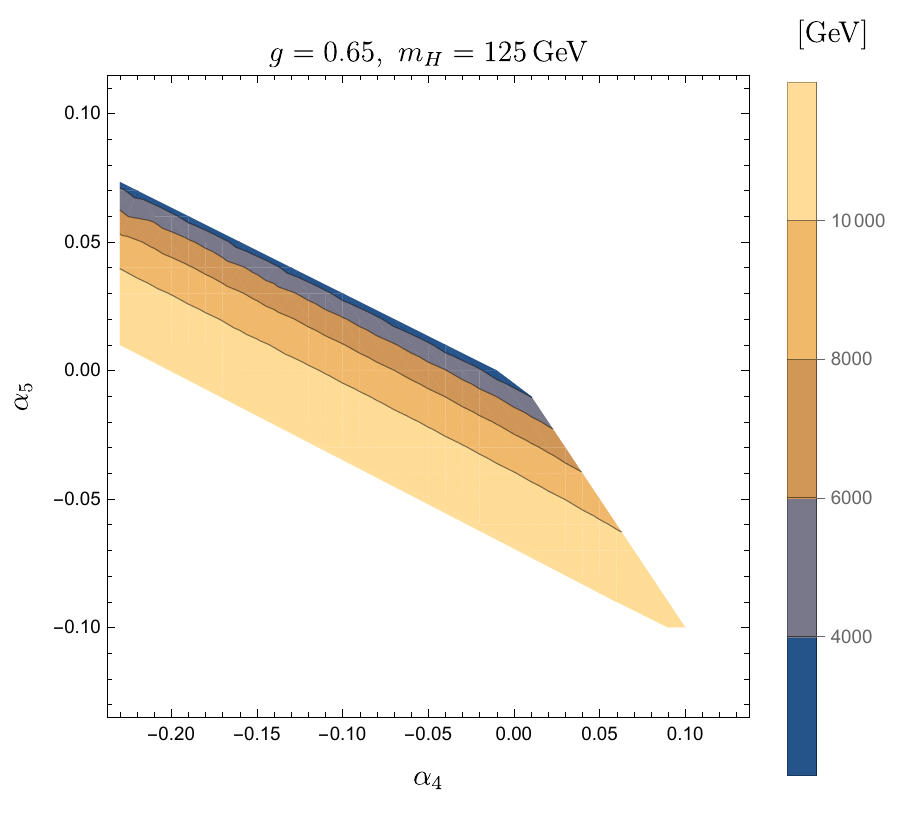} \hspace{0.em}
\includegraphics[width=0.49\textwidth]{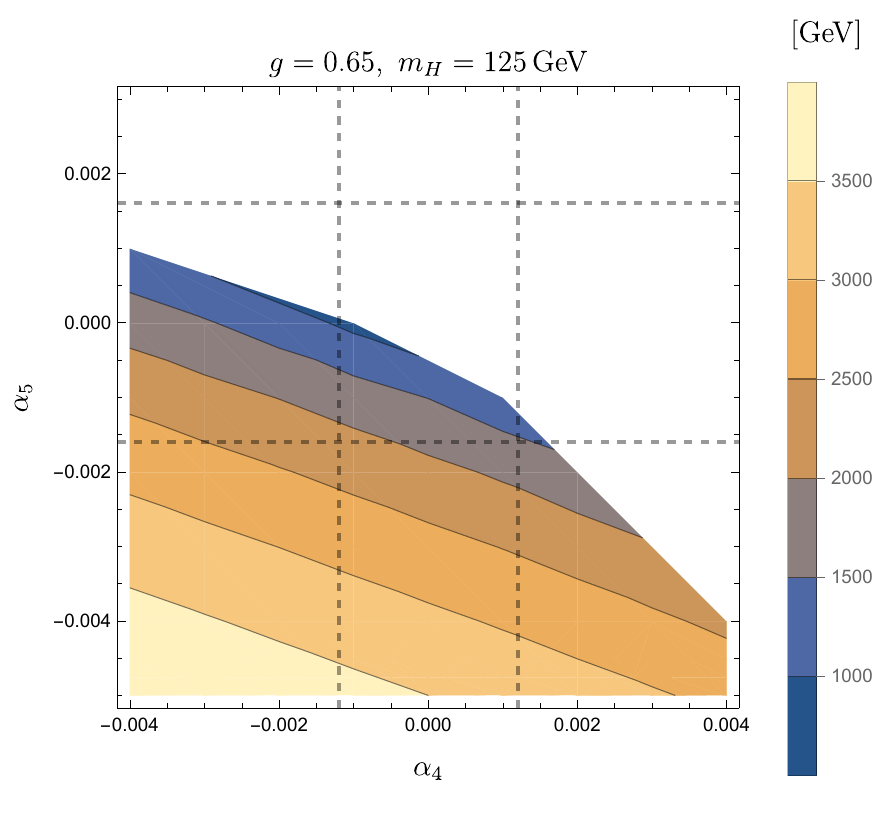} 
\caption{
Contour plot of the mass of the EW-Skyrmion in $\alpha_4$ - $\alpha_5$  plane.
White blank region represents that there is no stable solution.
The maximum value of the skyrmion mass is about $12\, \mr{TeV}$.
The right panel is an enlarged version of the left one.
The black dashed lines indicate the most stringent experimental bound from CMS analysis~\cite{CMS:2019qfk}.
See Sec.~\ref{sec:abundance} for more details.
}
\label{fig:energy-cont}
\end{figure}

\begin{figure}[tbp]
\centering
\includegraphics[width=0.47\textwidth]{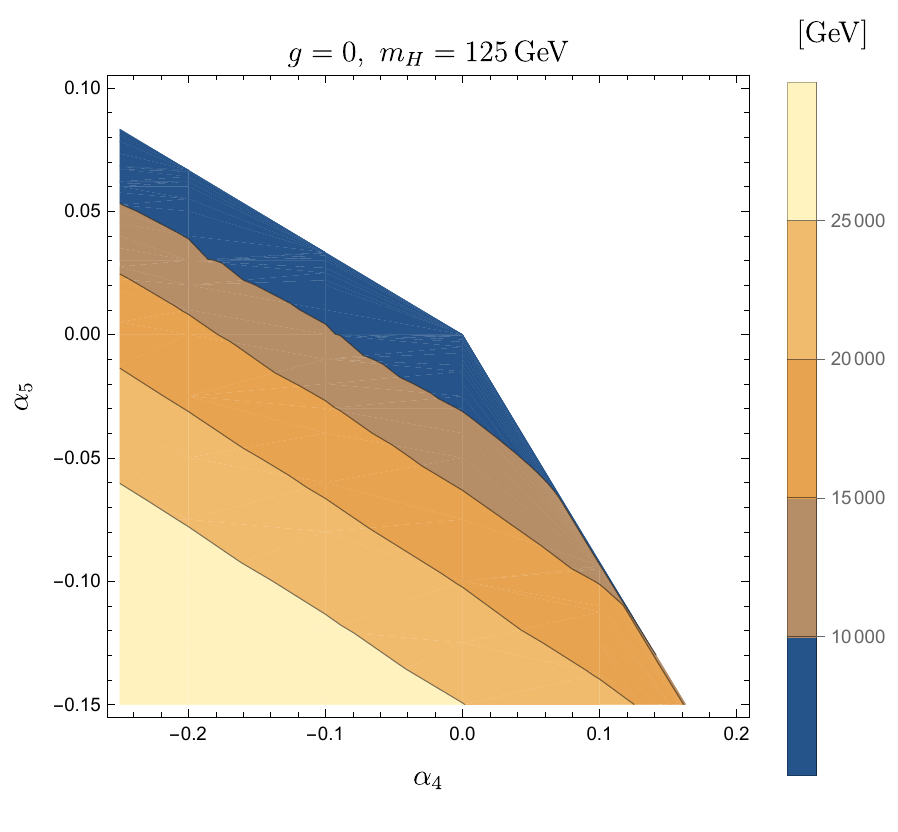}
\caption{
Contour plot of the mass of the EW-Skyrmion in $\alpha_4$ - $\alpha_5$  plane for $g=0$.
White blank region represents that there is no stable solution.
This is the same result presented in Ref.~\cite{Kitano:2017zqw}.
The mass of the skyrmion can be arbitrarily large, depending on $\alpha_4$ and $\alpha_5$.
}
\label{fig:energy-cont-g0}
\end{figure}

Finally, we also obtain solutions for more general cases 
by relaxing the condition $\alpha_4 = -\alpha_5 $, \textit{i.e.}, 
taking $\alpha_4$ and $\alpha_5 $ as independent values in Lagrangian Eq.~\eqref{eq:Lp4}.
Fig.~\ref{fig:energy-cont} shows a contour plot for the dependence of the energy on the two parameters $\alpha_4$ and $\alpha_5$
keeping $m_H $, $v_\mr{EW}$, and $g$ as the physical values
\begin{equation}
v_\mr{EW} = 246 ~ \mr{GeV},~ m_H = 125~ \mr{GeV}, ~ g=0.65\, .
\end{equation}
For comparison, we also present the previous result studied in Ref.~\cite{Kitano:2017zqw} in Fig.~\ref{fig:energy-cont-g0},
in which the gauge fields are decoupled by setting $g=0$.
The Higgs VEV and mass are taken as the physical values.
In both figures, white blank region represents that there is no stable solution.
We found that 
the EW-Skyrmion mass is bounded from above at around $12\, \mr{TeV}$
as shown in Fig.~\ref{fig:energy-cont}.
This is in contrast to the previous case without the gauge fields (Fig.~\ref{fig:energy-cont-g0}),
in which the energy can be arbitrarily large for large negative values of $\alpha_4$ and $\alpha_5$.
We conclude that the EW-Skyrmion exists as a stable soliton object even when the $SU(2)_W$ gauge fields are included in the Lagrangian 
for a certain parameter range of $\alpha_4$ and $\alpha_5$.
Experimentally allowed region for $\alpha_4$ and $\alpha_5$ will be presented in the next section.

Before closing this section, we should note that 
the EW-Skyrmion decays with a finite lifetime by quantum or thermal processes
even though it is a stable solution to the classical EOMs
because its field configuration can be continuously deformed into the vacuum as stated above.
Importantly, the EW-Skyrmion carries a non-trivial Chern-Simons number,
and hence its decay produces the baryon number via the chiral anomaly:
\begin{align}
\Delta (B+L) = 6 \Delta N_{CS} = 6 \Delta Q \,,\label{172957_4Jul21}
\end{align}
where $B$ and $L$ are the baryon and lepton number, respectively.
The second equality in Eq.~\eqref{172957_4Jul21} follows from the definition of $Q$, Eq.~\eqref{173008_4Jul21},
and the fact that $N_H$ is topological, \textit{i.e.} $\Delta N_H=0$.
While the decay width at the zero temperature is exponentially suppressed,
it is not negligible at the finite temperature as in the case of the sphaleron process.
This finite-temperature decay together with Eq.~\eqref{172957_4Jul21} makes the scenario possible that the EW-Skyrmion is an asymmetric dark matter,
that is,
if the (inverse) decay process is in the chemical equilibrium in the early universe, 
we can predict the relation between the DM relic abundance and the baryon number density,
as we will discuss in the next section.
It is interesting to note that $\Delta Q$ is non-integer when the EW-Skyrmion decays.
This means that $\Delta (B+L)$ is also non-integer 
while the final state should have an integer $B+L$.
For consistency, the EW-Skyrmion should carry a small non-integer $B+L$ 
originated from non-trivial configuration of the gauge fields with $N_{CS}\neq 0$
through the chiral anomaly.%
\footnote{We thank Ryutaro Matsudo and Kyohei Mukaida for discussion on this point.}

\section{Dark matter abundance}
\label{sec:abundance}

By assuming the geometric cross section, $\pi R^2$, for the pair annihilation of EW-Skyrmion and anti-EW-Skyrmion, 
the thermal relic abundance is estimated as
$ \Omega h^2 \sim 0.1\times \left( 60 \, \mr{GeV}/M\right)^2 \, ,$
which is too small as DM for TeV-mass EW-Skyrmions that are experimentally allowed~\cite{Kitano:2016ooc}.
Therefore, we here discuss the possibility of having primordial asymmetry of the EW-Skyrmions,
which remains as DM today.
We find that the scenario is consistent with both experimental constraints and the baryon asymmetry of the Universe.

\subsection{EW-Skyrmion as asymmetric dark matter}

We here propose a scenario that the EW-Skyrmion is an asymmetric dark matter.
As is shown in Eq.~\eqref{172957_4Jul21}, the decay of the EW-Skyrmion can produce the baryon number and vice versa,
and hence we can relate them once they get in thermal equilibrium.
A similar study in the context of the techni-hadron as an asymmetric dark matter candidate was done in Ref.~\cite{Barr:1990ca}.
See also Ref.~\cite{Harvey:1990qw}.
We hereafter call the decay process of the EW-Skyrmion ``sphaleron-like process''  
although it is different from the ordinary sphaleron process between two degenerated vacua.
We assume that the process becomes inactive at a temperature $T \simeq T^\ast$
and the baryon number and the DM number are frozen out for $T \ll T^\ast$.
Because we rely on the Higgs effective field theory approach,
$T^\ast$ should not be larger than the critical temperature $T_c$ of the EW phase transition
and is typically of the order of $10^2~\mr{GeV}$.

To discuss thermal history,
we start with a formula consisting of the number density and the chemical potential 
for a particle species labeled with $i$:
\begin{equation}
n_i(T) = C_i(m_i,T)\ \frac{\mu_i}{T},
\end{equation}
where 
\begin{align}
C_i(m_i,T) = g_i \, f(m_i/T) \, T^3
\end{align}
and $g_i$ is the number of the internal degrees of freedom.
The function $f(x)$ has different expressions depending on the statistics of the particles,
\begin{align}
f(x) = 
 \begin{cases}
    \frac{1}{4\pi^2} \int_0^\infty \frac{y^2 dy}{\cosh^2\left(\frac{1}{2}\sqrt{y^2+x^2}\right)} & {\rm (fermions)} \\
    \frac{1}{4\pi^2} \int_0^\infty \frac{y^2 dy}{\sinh^2\left(\frac{1}{2}\sqrt{y^2+x^2}\right)} & {\rm (bosons)}
 \end{cases} \, ,
\end{align}
and has asymptotic formulae as
\begin{align}
f(0) = 
 \begin{cases}
    1/6 & {\rm (fermions)} \\
    1/3 & {\rm (bosons)}
 \end{cases}
\end{align}
and
\begin{align}
f(x \gg 1) \ \sim \ 2 \left(\frac{x}{2\pi}\right)^{3/2} e^{-x}\ \ 
 \begin{cases}
    {\rm (fermions)} \\
    {\rm (bosons)}
 \end{cases}  \, .
\end{align}

\begin{table}[tb]
\centering
\begin{tabular}[tb]{|c|c|c|c|c|c|c|c|c|c|}
\hline
$W^- $&$\phi^0 $&$u_L $&$u_R $&$d_L $&$d_R $&$\nu_{Li} $&$e_{Li} $&$e_{Ri}$ \\ \hline
$\mu_{W} $&$\mu_{0} $&$\mu_{uL}$&$\mu_{uR}$&$\mu_{dL}$&$\mu_{dR}$&$\mu_{i} $&$\mu_{iL}$&$\mu_{iR}$ \\ \hline
\end{tabular}
\caption{
Table for the definition of chemical potentials in the EW broken phase.
$\phi^0$ is the neutral component of the Higgs doublet, \textit{i.e.}, the physical Higgs boson.
The indices $R$ and $L$ represent chiralities of the fermions.
$u_{R(L)}$ and $d_{R(L)}$ denote the up-type and down-type quarks, respectively.
They are assumed to be independent of the generations because of the gluon interaction.
$e_{R(L)i}$ and $\nu_{Li}$ are the charged leptons and the left-handed neutrinos
with $i$ being the label for the generations.
There are $6+3N$ chemical potentials in total.
}
\label{184224_4Jul21}
\end{table}

Then, we define $6+3N$ chemical potentials as Table~\ref{184224_4Jul21},
where $N$ is the number of the fermion generations ($N=3$ in the SM).
Using these chemical potentials, we express the baryon and lepton number density as
\begin{align}
\frac{n_B}{f(0)_{\rm fermion}T^2} &= N(\mu_{uL}+\mu_{uR})+N(\mu_{dL}+\mu_{dR}) \nonumber\\
&=4N\mu_{uL} + 2N\mu_W
\label{eq:nB}
\end{align}
\begin{align}
\frac{n_L}{f(0)_{\rm fermion}T^2} &= \sum_{i=1\sim N} (\mu_i+\mu_{iL}+\mu_{iR})\nonumber\\
&= 3\mu + 2N\mu_W  - N\mu_0
\end{align}
where we have assumed that the masses of the fermions are negligible
and $\mu \equiv \sum_{i=1 \sim N} \mu_i$.
On the other hand,
the number density of the EW-Skyrmion, \textit{i.e.,} the DM number density is given as
\begin{align}
\frac{n_{DM}}{f\left(m_{DM}(T)/T\right)\, T^2}= \mu_{DM} 
\end{align}
with its chemical potential $\mu_{DM}$.

We assume that the electroweak and the Yukawa interactions are in the chemical equilibrium for $T\simeq T^\ast$,
and hence obtain the following relations:
\begin{align}
\mu_{dL}&= \mu_{uL}+\mu_W \ \ \ \  (W^- \leftrightarrow \bar{u}_L + d_L)\\
\mu_{iL} &= \mu_{i}+\mu_W \ \ \ \  (W^- \leftrightarrow \bar{\nu}_{iL} + e_{iL})\\
\mu_{uR}&= \mu_{0}+\mu_{uL} \ \ \ \  (\phi^0 \leftrightarrow \bar{u}_L + u_R)\\
\mu_{dR} &= -\mu_{0}+\mu_W + \mu_{uL} \ \ \ \  (\phi^0 \leftrightarrow d_L + \bar{d}_R)\\
\mu_{iR} &= -\mu_{0}+\mu_W+\mu_i \ \ \ \  (\phi^0 \leftrightarrow e_{iL} + \bar{e}_{iR}) \, .
\end{align}
By using these $3+2N$ relations, we can express all of the chemical potentials in terms of $3+N$ chemical potentials: $\mu_W$, $\mu_0$, $\mu_{uL}$, and $\mu_i$.
Note that $\mu_0=0$ because of the condensation of $\phi^0$.
In addition, due to the charge conservation, 
the charge neutrality condition must be met,
\begin{align}
&2N(\mu_{uL}+\mu_{uR}) -N(\mu_{dL}+\mu_{dR}) -\sum_{i} (\mu_{iL}+\mu_{iR}) -6\mu_W \nonumber\\
&= 2N\mu_{uL} -2\mu -(4N+6)\mu_W +(4N+2)\mu_0 \nonumber\\
&=0 \,.
\label{eq:charge}
\end{align}
%
%

Because the sphaleron-like process is in the chemical equilibrium at $T\simeq T^\ast$,
we obtain the relation between the chemical potentials of the baryon, lepton, and the dark matter (EW-Skyrmion).
From Eq.~\eqref{172957_4Jul21}, 
a single dark matter decays into fermions with $B+L=2 N $.
Then we have the following relation \cite{Barr:1990ca}:
\begin{align}
N(\mu_{uL}+2\mu_{dL}) + \sum_i \mu_i + \mu_{DM} = 0 \, ,
\end{align}
or equivalently,
\begin{align}
3N\mu_{uL}+2N\mu_W  + \mu + \mu_{DM} = 0
\label{eq:sph} \,.
\end{align}

\begin{figure}[tbp]
\begin{center}
\includegraphics[width=0.7\textwidth]{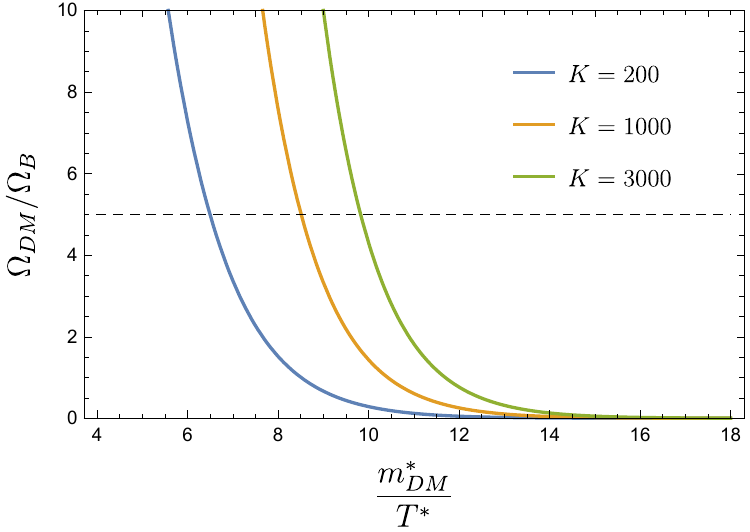}
\caption{
Plot of $\Omega_{DM}/\Omega_{B}$ versus $m^\ast_{DM}/T^\ast$ for several choices of $K$.
The black dashed line indicates an appropriate value to explain the observed ratio $\Omega_{DM}/\Omega_{B}\simeq 5$.
The parameter $K$ depends on the thermal history before the EW phase transition and is treated as a free parameter.
}
\label{Fig:abundance}
\end{center}
\end{figure}

By using Eqs.~\eqref{eq:nB}-\eqref{eq:sph}, 
we obtain the following relation:
\begin{align}
\frac{n_{DM}}{n_B} = - \frac{f\left(m^\ast_{DM}/T^\ast\right)}{f(0)_{\rm fermion}}
\left[ \frac{16N+27}{22N+36} + \frac{4N+6}{11N+18}\frac{n_L}{n_B}  \right] \, .
\end{align}
Here, $m^\ast_{DM}$ represents the dark matter mass at $T\simeq T^\ast$.
 By substituting $N=3$, we obtain
\begin{align}
\frac{n_{DM}}{n_B} = - \frac{f\left(m^\ast_{DM}/T^\ast\right)}{f(0)_{\rm fermion}}
\left[ \frac{25}{34} + \frac{6}{17}\frac{n_L}{n_B}  \right]  \,.
\end{align}
Then the ratio of the present relic abundances of the dark matter and the baryon is expressed as
\begin{align}
\frac{\Omega_{DM}}{\Omega_{B}} = \frac{n_{DM}}{n_B}\frac{m_{DM}}{m_p} &= - \frac{f\left(m^\ast_{DM}/T^\ast\right)}{f(0)_{\rm fermion}}
\left[ \frac{25}{34} + \frac{6}{17}\frac{n_L}{n_B}  \right] \frac{m_{DM}}{m_p}\label{180142_10Jul21} \\
&\sim K \left(\frac{m^\ast_{DM}}{T^\ast}\right)^{3/2} \exp{\left(-\frac{m^\ast_{DM}}{T^\ast}\right)}
\end{align}
where
\begin{align}
K &\equiv \frac{6}{\sqrt{2}\pi^{3/2}} 
\left[ \frac{25}{34} + \frac{6}{17}\frac{n_L}{n_B}  \right] \frac{m_{DM}}{m_p}  
\end{align}
and $m_p$ is the proton mass.
Fig.~\ref{Fig:abundance} shows the plot of $\Omega_{DM}/\Omega_{B}$ versus $m^\ast_{DM}/T^\ast$.
Note that $K$ itself is not determined by the present argument
since it depends on $n_L/n_B$.
For typical values of $m_{DM} \sim m_{DM}^* \sim {\rm TeV}$, 
$T^\ast \sim 100~\mr {GeV}$ and $n_L/n_B \sim O(1)$, 
we find that the desired relation, $\Omega_{DM} / \Omega_B \sim O(1)$ can be realized.

Instead of using $n_L/n_B$,
it is convenient to rewrite Eq.~\eqref{180142_10Jul21} in terms of conserved quantities, $Y_{DM-B/3}\equiv (n_{DM} - n_B /3)/s$ and $Y_{B-L}\equiv (n_{B}- n_L)/s$ with $s$ being the entropy density,
 as
\begin{align}
 \frac{\Omega_{DM}}{\Omega_{B}} 
%
&=X \frac{ 111 \, Y_{DM-B/3} + 12 \, Y_{B-L}}{-102 \, Y_{DM-B/3} + 36 X Y_{B-L}}\frac{m_{DM}}{m_p}\label{eq:Omega_ratio} 
\end{align}
with
\begin{equation}
X\equiv \frac{f\left(m^\ast_{DM}/T^\ast\right)}{f(0)_{\rm fermion}} \, .
\end{equation}
In this formula, since $Y_{DM-B/3}$ and $Y_{B-L}$ are conserved, one can use the values at which
they are produced.
While $X$ is very small, such as $\mathcal{O}(10^{-(4-5)})$ for $m_{DM}\sim \mathcal{O}(1) \, \mr{TeV}$,
a large factor $m_{DM}/m_p$ can compensate such a suppression, 
and one can obtain $\mathcal{O}(1)$ values of $\Omega_{DM}/\Omega_B$ when $Y_{DM-B/3}\neq 0$.
On the other hand, for $n_{DM} -n_B/3= 0$ and $n_{B-L}=-n_L \neq 0$ 
as in the case of the leptogenesis scenario,
the $X$-dependence is canceled out in Eq.~\eqref{eq:Omega_ratio},
and one obtains too large ratio
$\Omega_{DM}/\Omega_{B} = m_{DM}/3m_p  \gg 1$.
For successful cosmology, the baryogenesis requires to generate $n_{DM} -n_B/3$.

Note that, before the EW phase transition,
the EW-Skyrmion itself does not exist since there is no NG mode from the Higgs field.
Nevertheless, the corresponding quantum number can be defined
in the language of the UV model behind the Higgs sector.
Thus it is possible to generate $n_{DM}$ before the EW phase transition by the dynamics in the UV model.

\subsection{Experimental constraints}

As we mentioned in the previous section, $\alpha_4$ and $\alpha_5$ parameterize the deviation from the SM value of quartic gauge couplings, 
and magnitude of them are experimentally constrained by measurements of weak gauge boson scattering amplitudes. 
The currently available constraints from ATLAS and CMS analyses of LHC experiment are given as
\begin{equation}
\text{(CMS)\cite{CMS:2019qfk}} \hspace{1em} |\alpha_4| \lesssim 0.0012, \hspace{1em} |\alpha_5| \lesssim 0.0016
\end{equation}
\begin{equation}
\text{(ATLAS)\cite{Aaboud:2016uuk}} \hspace{1em} -0.024 < \alpha_4 < 0.030, \hspace{1em} -0.028 < \alpha_5 < 0.033
\end{equation}
with 95 \% CL limits.
Here we have translated the CMS result analyzed in the linear representation of the Higgs field
into the non-linear one.
See Appendix \ref{183447_10Jul21} for details.
These bounds are obtained by using the semileptonic final states.
Combining these constraints and Fig.~\ref{fig:energy-cont},
an upper limit for the mass of the EW-Skyrmion is obtained as
\begin{equation}
 m_{DM} \lesssim 2.2 \, \mathrm{TeV} \,.
\end{equation}

On the other hand, there are stringent bounds from DM direct detection experiments,
which constrain the coupling to the nuclei for sub-TeV DM.
To discuss it, we assume the effective Higgs-portal-like coupling 
$\mathcal{L}_{eff.}\simeq -\kappa |S|^2 |H|^2$,
where $S$ and $H$ denote an EW-Skyrmion field and the SM Higgs doublet, respectively.
Then we obtain the spin-independent cross section for the elastic scattering between the EW-Skyrmion and the nucleon as
\begin{align}
 \sigma_\mr{SI} &\simeq \frac{\kappa^2 m_N^4 f^2}{\pi m_{DM}^2m_h^4}\\
& \simeq  \left(\frac{\kappa}{0.1} \right)^2 \left(\frac{1 \, \mr{TeV}}{m_{DM}} \right)^2 \left(\frac{f}{0.3} \right)^2 \times 3.6 \times 10^{-46}~ \mr{cm}^2 \, ,
\end{align}
where $m_N$ is the nucleon mass and $f$ is the form factor, which is taken to be $0.3$ here.
The most stringent upper bound on $\sigma_\mr{SI}$ is obtained by the PandaX-4T and XENON1T experiments \cite{XENON:2017vdw,XENON:2018voc,PandaX-4T:2021bab}.
When we take $\kappa = 0.1$, the bound is translated to the lower bound of the EW-Skyrmion mass as 
\begin{equation}
m_{DM} \gtrsim 0.9 \, \mr{TeV} \, .
\end{equation}
This value strongly depends on the choice of $\kappa$ and becomes $2.6 \, \mr{TeV}$ and $0.6 \, \mr{TeV}$ for $\kappa = 0.5$ and $\kappa=0.05$, respectively.
Therefore,  we conclude that the EW-Skyrmion with mass between $0.9 \, \mr{TeV}$ and $2.2 \, \mr{TeV}$, 
with certain amount of uncertainty, 
is consistent with the current experimental bounds, while explaining the ratio of the relic abundances of dark matter and the baryon number in the universe.

\section{Summary}
\label{sec:summary}
We have proposed a scenario that the EW-Skyrmion,
a meta-stable soliton consisting of the Higgs and the EW gauge fields,
plays a role of the asymmetric DM.
In contrast to the case that the gauge fields are ignored,
the EW-Skyrmion is not topologically stable but classically 
for some parameter region of the magnitudes of the higher derivative terms.
At high temperatures, the EW-Skyrmion can decay 
through a process which changes the Chern-Simons number $N_{CS}$.
Such a process generates the baryon number due to the $B+L$ anomaly,
which enables us to relate the baryon number density and the DM number density.
From the experimental results to look for the anomalous quartic gauge coupling,
we have imposed the experimental bounds on the magnitudes of the higher-derivative terms, $\alpha_4$ and $\alpha_5$,
which are translated into the upper limit of the mass of the EW-Skyrmion.
In addition, we have studied the experimental bound from the DM direct detection in XENON1T experiment.
We have obtained a lower bound for the mass with uncertainty.
We conclude that the scenario of the EW-Skyrmion as the asymmetric DM can naturally explain $\Omega_{DM}/\Omega_B \simeq 5$
within the experimental bounds of the mass of the EW-Skyrmion, $0.8 \, \mr{TeV}\lesssim m_{DM} \lesssim 2.2 \, \mr{TeV}$.
We emphasize that our scenario is independent of UV physics because we have utilized the effective Lagrangian for the Higgs field.

Let us comment on bounds on our scenario from indirect DM searches using cosmic rays.
Since the DM relic abundance is dominated by the asymmetric part,
the pair annihilation of the EW-Skyrmion and anti EW-Skyrmion is suppressed to occur in the present universe.
Therefore indirect DM searches give no severe constraints.

In this paper, we have ignored the $U(1)_Y$ coupling, $g'=0$, while switching $g$ on.
This coupling is expected to shift the mass of the EW-Skyrmion by a few percent level
and not to change the present argument qualitatively.
We should confirm that this is true by numerical analysis and 
that there is still allowed parameter space,
which will be done elsewhere.
In addition, we have expressed the temperature
in which the sphaleron-like process is out of equilibrium as $T^\ast$.
In principle, this is calculable and determined by the energy of a sphaleron-like solution,
which is sitting at the top of the energy barrier between the EW-Skyrmion and the vacuum configurations.
This would also be theoretically interesting in the context of unstable classical solutions in the gauged non-linear sigma model.

For the future perspective, 
improvements of the experiments, 
XENON experiment and the collider experiments at (HL-)LHC and/or ILC,
will impose more stringent constraints on our scenario.
In particular, the bounds of $\alpha_4$ and $\alpha_5$ obtained from measuring anomalous quartic gauge coupling are very crucial.
If any deviation of the parameters from the SM value ($\alpha_4 = \alpha_5=0$) will be detected by such precision measurements,
it strongly supports our scenario and can predict the DM mass. 
In addition, since leptogenesis is incompatible with our scenario as stated above, 
it gives some implications on UV models beyond the SM.

\acknowledgments
We thank Ryutaro Matsudo, Kyohei Mukaida, and Muneto Nitta for useful discussions. 
The work is supported by JSPS KAKENHI Grant Nos. JP19H00689 (RK) and JP21J01117 (YH), and MEXT KAKENHI Grant No. JP18H05542 (RK).

\appendix

\section{Perturbative analysis of (in)stability}
\label{App:instability}
We show based on a perturbative analysis that the stable EW-Skyrmion is not expected for large $g^2 \alpha$.
To this end, it is convenient to clarify what terms of the energy functional have potentials leading to the instability.
For the sake of convenience for the reader,
we present the energy functional again:
\begin{align}
 \frac{E}{4\pi}&= \int _0 ^\infty dr ~ \Big[ \frac{1}{2 g^2 r^2}
\left[2(r^2 \delta^2 -1) \chi^2 + 2 r^2 \chi'^2 +\chi^4 +1\right] \nonumber  \\
&+ \frac{1}{8}(\delta^2 - 4 \delta\theta')\left[r^2 v^2 (1+\phi)^2 + 8 \alpha (1+\chi^2 -2 \chi \cos 2\theta)\right] \nonumber \\
&+ \frac{1}{4} \lambda r^2 v^4 (\phi^4 + 4 \phi^3 + 4 \phi^2)
+ \frac{1}{2}r^2 v^2 (\theta'^2 + \phi'^2)
+\frac{1}{4} (v^2 + 16 \alpha\theta'^2)(1+\chi^2 -2 \chi \cos 2\theta)  \nonumber\\
&+\frac{1}{2 r^2}\alpha (1+\chi^4 -4 \chi^3 \cos 2\theta -4\chi \cos 2\theta + 2\chi^2\cos4\theta + 4\chi^2 ) \nonumber \\
&+ \frac{1}{4}v^2 (\phi^2 + 2\phi) (2r^2 \theta'^2 - 2\chi \cos 2\theta + \chi^2 +1) \Big] \, , \label{170702_17Jul21}
\end{align}
which is the same as Eq.~\eqref{185321_22May21}.
We have assumed $\alpha_4=-\alpha_5=\alpha$ for simplicity.
As is stated above, $\delta$ does not have a kinetic term, which means that it is an auxiliary field and is explicitly solvable.
The solved value is 
\begin{equation}
 \bar{\delta} \equiv 2 \theta' 
\left(1- \frac{8 \chi^2}{g^2}
\left[r^2 v^2 (1+\phi)^2 + 8 \alpha (1 + \chi^2 -2 \chi \cos 2\theta) + 8 \chi ^2 /g^2\right]^{-1}
\right) .
\end{equation}
Note that, when we set $\delta = \bar \delta$ and $\chi \to 0$, 
all terms including $\theta(r)$ in Eq.~\eqref{170702_17Jul21} vanish.
Thus a fluctuation around $\theta(r)$ does not change the energy, \textit{i.e.}, it is a flat direction.
In general, such a criticality probes the emergence of the instability.
Indeed, if $\chi$ passes through 0 and goes to negative values,
the instability for $\theta$ appears
because, using $\delta = \bar \delta$, the only quadratic term of $\theta$ comes from
\begin{equation}
-2\chi \cos2\theta \simeq -2 \chi + 4\chi \theta^2 + \mathcal{O}(\theta^4),
\end{equation}
which is clearly tachyonic for $\chi < 0$.

Now, the problem has reduced to the following question: does $\chi(r)$ vanish (or become negative) at some point for large $g^2 \alpha$?
To see this, we recall that the gauge field $W_i$ is generated through the linearized EOM \eqref{111931_24May21} 
(see Sec.~\ref{sec:gauge}) at the leading order of $g^2 \alpha$.
Substituting the ansatz of $U$ and $h$, Eqs.~\eqref{144403_10May21} and \eqref{201903_3Jul21}, 
we can rewrite the $SU(2)_W$ current as
\begin{align}
 J_i^a   \,
=& - g v_\mathrm{EW}^2 (1+ \phi)^2
 \sin^2 \theta(r) \frac{\epsilon^{aib}\hat{x}_b}{r} \nonumber \\
&-8 g \alpha \left[
 ( \cos 2\theta\, \theta')^2-\frac{\sin^2\theta}{r^2}
\right] 
 \sin^2 \theta(r) \frac{\epsilon^{aib}\hat{x}_b}{r} + (\cdots)
\end{align}
where we have omitted terms that include symmetric tensors with respect to the indices $i$ and $a$
because they are not relevant for $\chi$.
Then, projecting onto the anti symmetric part with respect to the indices $i$ and $a$ in the linearized EOM,
we can solve it for $\chi$ as
\begin{equation}
\partial^2 \left[\frac{1- \chi}{gr} \epsilon^{aib}\hat{x}_b\right] \propto g \alpha R^{-2} \sin^2 \theta(r) \frac{\epsilon^{aib}\hat{x}_b}{r}
\end{equation}
where we have replaced $v_\mathrm{EW}^2$ by $\alpha^{-1}R^{-2}$ and dropped $\mathcal{O}(1)$ factors in $J_i^a$.
Roughly, the derivative acting on $W_i$ can be replaced by $R^{-1}$, and we obtain
\begin{align}
1-\chi \propto  g^2 \alpha
 \sin^2 \theta(r) \, .
\end{align}
Hence $\chi$ significantly deviates from unity when $g^2 \alpha$ is large.
Therefore, within the leading order of the perturbative analysis, 
$\chi$ is expected for sufficiently large $g^2 \alpha$ to vanish at some points with $\sin \theta(r) \neq 0,$
which generates the instability.
This behavior of the vanishing $\chi$ and the emergence of the instability is confirmed by numerical analysis.

\section{ATLAS and CMS experimental constraints for aQGC}
\label{183447_10Jul21}

In ATLAS analysis (\textit{e.g.}, Refs.~\cite{Aaboud:2016ffv,Aaboud:2016uuk}), 
operators described in the non-linear representation are used to parameterize the anomalous quartic gauge couplings (aQGCs).
Among them, we focus on two operators 
\begin{equation}
 \alpha_4\mr{Tr} \left[D_\mu U^\dagger D_\nu U \right] \mr{Tr} \left[D^\mu U^\dagger D^\nu U \right]
\end{equation}
\begin{equation}
 \alpha_5\mr{Tr} \left[D_\mu U^\dagger D^\mu U \right] \mr{Tr} \left[D_\nu U^\dagger D^\nu U \right] \,
\end{equation}
which are nothing but the higher derivative terms $\mathcal{L}_{p^4}$ Eq.~\eqref{eq:Lp4} we use in this paper.

On the other hand, in CMS analysis, the linear representation for the Higgs field is used (\textit{e.g.}, Refs.~\cite{CMS:2019qfk,CMS:2020fqz} ).
Operators that give the same scattering process with the $\alpha_4$ and $\alpha_5$ terms in the non-linear representation are
\begin{equation}
 \frac{f_0}{\Lambda^4} \left[(D_\mu \Phi)^\dagger D_\nu \Phi\right]\left[(D^\mu \Phi)^\dagger D^\nu \Phi\right]
\end{equation}
\begin{equation}
 \frac{f_1}{\Lambda^4} \left[(D_\mu \Phi)^\dagger D^\nu \Phi\right]\left[(D_\nu \Phi)^\dagger D^\nu \Phi\right] \, .
\end{equation}
Following Ref.~\cite{Eboli:2006wa},
we can relate them by the following equations
\begin{equation}
 \frac{f_0}{\Lambda^4} \Leftrightarrow \frac{8\alpha_4}{v_\mr{EW}^4}, \hspace{2em}  \frac{f_1}{\Lambda^4} \Leftrightarrow \frac{8\alpha_5}{v_\mr{EW}^4} \, .
\end{equation}
Using these, we can translate bounds on $f_0/\Lambda^4$ and $f_1/\Lambda^4$ into those on $\alpha_4$ and $\alpha_5$.
The most stringent bound on the aQGCs is given in Ref.~\cite{CMS:2019qfk} by CMS as
\begin{equation}
  \left|\frac{f_0}{\Lambda^4} \right| \leq 2.7 ~  (\mr{TeV}^{-4}) , \h{2em}
  -3.3 ~(\mr{TeV}^{-4})\leq \frac{f_1}{\Lambda^4} \leq 3.4 ~ (\mr{TeV}^{-4})  ,
\end{equation}
which are translated to 
\begin{equation}
|\alpha_4| \lesssim 0.0012, \hspace{2em} |\alpha_5| \lesssim 0.0016.
\end{equation}

\bibliographystyle{jhep}
\bibliography{./references}

\end{document}